\documentclass{PoS}

\title{Putting M theory on a computer}

\ShortTitle{Putting M theory on a computer}

\author{\speaker{Jun Nishimura}\\
High Energy Accelerator Research Organization (KEK), 
		Tsukuba 305-0801, Japan,\\
and Graduate University for Advanced Studies (SOKENDAI),
Tsukuba 305-0801, Japan\\
        E-mail: \email{jnishi@post.kek.jp}}

\author{Konstantinos N. Anagnostopoulos\\
National Technical University of Athens,
Zografou Campus, GR-15780 Athens, Greece
\\
        E-mail: \email{konstant@mail.ntua.gr}}

\author{Masanori Hanada\\
Theoretical Physics Laboratory, 
RIKEN Nishina Center,\\
2-1 Hirosawa, Wako, Saitama 351-0198, Japan
\\
        E-mail: \email{hana@riken.jp}}

\author{Shingo Takeuchi\\
Graduate University for Advanced Studies (SOKENDAI),
Tsukuba 305-0801, Japan
\\
        E-mail: \email{shingo@post.kek.jp}}

\abstract{
We propose a non-lattice
simulation
for studying supersymmetric
matrix quantum mechanics in a non-perturbative manner. 
In particular, our method enables us to put M theory on a
computer 
based on its matrix formulation proposed by Banks, 
Fischler, Shenker and Susskind. 
Here we present Monte Carlo results of the same matrix model
but in a different parameter region, which corresponds
to the 't Hooft large-$N$ limit at finite temperature.
In the strong coupling limit
the model has a dual description in terms
of the $N$ D0-brane solution in 10d type IIA supergravity.
Our results provide highly nontrivial evidences for the
conjectured duality.
In particular, the energy (and hence the entropy) of the
non-extremal black hole has been reproduced by solving directly
the strongly coupled dynamics of the D0-brane effective theory.
}

\FullConference{The XXV International Symposium on Lattice Field Theory\\
		 July 30 - August 4 2007\\
		 Regensburg, Germany}

\newcommand {\beq} {\begin{equation}}
\newcommand {\eeq} {\end{equation}}
\newcommand {\beqa}{\begin{eqnarray}}
\newcommand {\eeqa}{\end{eqnarray}}
\newcommand {\del} {\partial}
\newcommand {\tr}{{\rm tr\,}}

\newcommand {\ee}{\mbox{e}}

\begin{document}

\section{Introduction}

Large-$N$ gauge theories
are playing more and more important
roles in theoretical particle physics.
In particular, they are considered to be 
useful in formulating
superstring/M theories non-perturbatively
extending the idea of matrix models,
which was successful 
for string theories in sub-critical dimensions.
For instance, it has been conjectured that
critical string/M theories
can be formulated in terms of matrix models, 
which can be formally obtained
by dimensionally reducing 
U($N$) super Yang-Mills theory 
in ten dimensions
to $D=0,1,2$ dimensions.
The $D=1$ case
corresponds to
the Matrix theory \cite{BFSS},
which is conjectured to describe
M Theory 
microscopically.

Another important conjecture, which has
been studied intensively over the decade, 
concerns the duality
between the strongly coupled
large-$N$ gauge theory and
the weakly coupled supergravity.
The best understood example is the
AdS/CFT correspondence,
but there are numerous extensions
to non-conformal field theories as well.
In particular, large-$N$ gauge theories
in low dimensions have been studied
intensively at finite temperature,
which revealed
intriguing connections
to the black-hole thermodynamics 
\cite{Itzhaki:1998dd,
Barbon:1998cr,%
KLL,
Aharony4}.

Monte Carlo simulation 
of large-$N$ gauge theories
is expected to
be very useful in order to
confirm these conjectures
or to make use of them.
Indeed, the totally reduced
models \cite{IKKT} (the gauge theory
reduced to $D=0$ dimension) 
have been studied in refs.\ 
\cite{red-sim,Anagnostopoulos:2001yb}.
In the $D\ge 1$ case, 
some sort of 
``discretization''
is needed in order to put the theory
on a computer. 
However, lattice simulation of supersymmetric gauge theories
is not straightforward.
In some cases 
the lack of manifest supersymmetry just 
necessitates cumbersome fine-tuning, but 
in the worse cases the chiral and/or 
Majorana nature of fermions makes it difficult to 
even formulate an appropriate lattice theory.
We propose to circumvent all these problems
inherent in the lattice approach
by adopting a {\em non-lattice} approach
\cite{Hanada-Nishimura-Takeuchi}
for one-dimensional 
supersymmetric gauge theories.
This approach, in particular, enables us to 
put M theory on a computer
using the Matrix theory \cite{BFSS},
which takes the form of
a 1d U($N$) gauge theory with 16 supercharges.

Here we demonstrate our approach
by studying the same model but in a different
parameter region, 
which corresponds
to the 't Hooft large-$N$ limit at finite temperature \cite{AHNT}.
In the strong coupling limit
the model has a dual description 
\cite{Itzhaki:1998dd}
in terms of the $N$ D0-brane solution 
in type IIA supergravity.
Our results provide highly nontrivial evidences for the
conjectured duality.
In particular, the energy (and hence the entropy) 
of the
non-extremal black hole has been reproduced by solving directly
the strongly coupled dynamics of the D0-brane effective theory.

\section{SUSY matrix quantum mechanics with 16 supercharges}

The model 
can be obtained formally by dimensionally
reducing 10d $\mathcal{N}=1$ 
super Yang-Mills theory to 1d.
The action is given by
\beq
S
= 
\frac{1}{g^2} \int_0^{\beta}  
d t \, 
\tr 
\bigg\{ 
\frac{1}{2} (D_t X_i)^2 - 
\frac{1}{4} [X_i , X_j]^2  
+ \frac{1}{2} \psi_\alpha D_t \psi_\alpha
- \frac{1}{2} \psi_\alpha (\gamma_i)_{\alpha\beta} 
 [X_i , \psi_\beta ]
\bigg\} \ ,
\label{cQM}
\eeq
where $D_t  = \del_t
  - i \, [A(t), \ \cdot \ ]$ represents the covariant derivative
with the gauge field $A(t)$ being an $N\times N$ Hermitian matrix.
This model
can be viewed as 
a one-dimensional U($N$) gauge theory with adjoint matters.
The bosonic matrices $X_i(t)$  $(i=1,\cdots,9)$
come from spatial components of the 10d gauge field,
while the fermionic matrices $\psi_\alpha(t)$
$(\alpha=1,\cdots , 16)$ come from
a Majorana-Weyl spinor in 10d.
The $16\times 16$ matrices $\gamma_i$
in (\ref{cQM}) act on spinor indices and
satisfy the Euclidean Clifford algebra
$\{ \gamma_i,\gamma_j \}= 2\delta_{ij}$.
We impose periodic and anti-periodic
boundary conditions
on the bosons and fermions, respectively.
The extent $\beta$ in the Euclidean time 
direction then corresponds to the inverse
temperature $\beta \equiv 1/T$.
The 't Hooft coupling constant
is given by $\lambda\equiv g^2 N$, and
the dimensionless effective coupling
constant is given by
$\tilde{\lambda}=\lambda/T^3$.
Without loss of generality we set $\lambda=1$,
hence low (high) $T$ corresponds to strong (weak)
coupling strength, respectively.

\section{Non-lattice simulation for SUSY matrix quantum mechanics}

We fix the gauge by the static diagonal gauge
\beq
A(t) = \frac{1}{\beta} {\rm diag} 
(\alpha_1 , \cdots \alpha_N) \ ,
\eeq
where $\alpha_a$ can be chosen to 
satisfy
the constraint 
$\max_a (\alpha_a) - \min_a (\alpha_a) 
\le 2\pi$
using the large gauge transformation.
We have to add to the action a term
$S_{\rm FP} =
- \sum_{a<b} 2 \ln 
\left| \sin \frac{\alpha_a - \alpha_b}{2}
\right|$,
which appears from the Faddeev-Popov procedure.

We make a Fourier expansion 
\beq
X_i ^{ab} (t) = \sum_{n=-\Lambda}^{\Lambda} 
\tilde{X}_{i n}^{ab} \ee^{i \omega n t}
 \ ; \
\psi_\alpha ^{ab} (t) = \sum_{r=-\Lambda'}^{\Lambda'}
\tilde{\psi}_{\alpha r}^{ab} \ee^{i \omega r t}
 \ ,
\eeq
%
where
$\omega = \frac{2 \pi }{\beta}$ and 
$\Lambda ' \equiv \Lambda-1/2$.
The indices $n$ and $r$ take integer and
half-integer values, respectively,
corresponding to the imposed 
boundary conditions.
Introducing a shorthand
notation
%
%
\beq
\Bigl(f^{(1)}  \cdots  f^{(p)}\Bigr)_n 
\equiv \sum_{k_1 + \cdots + k_{p}=n}
f^{(1)}_{k_1} \cdots f^{(p)}_{k_p} \ ,
\eeq
we can write the action 
(\ref{cQM}) as
$S=S_{\rm b}+S_{\rm f}$, where
\beqa
S_{\rm b}
&=&  N \beta
\Bigg[
\frac{1}{2} \sum_{n=-\Lambda}^{\Lambda} 
\left( n \omega - \frac{\alpha_a - \alpha_b}{\beta} 
\right)
^2   \tilde{X}_{i , -n}^{ba} \tilde{X}_{i n}^{ab}
- \frac{1}{4} 
 \tr \Bigl( [ \tilde{X}_{i} , \tilde{X}_{j}]^2  \Bigr)_0
\Bigg] 
\nonumber \\
S_{\rm f}
&=& \frac{1}{2}
N \beta \sum_{r=-\Lambda '}^{\Lambda '} \Biggl[
i 
\left(
r \omega - \frac{\alpha_a - \alpha_b}{\beta} 
\right)
\tilde{\psi}_{\alpha , -r}^{ba} \tilde{\psi}_{\alpha r}^{ab} 
 - (\gamma_i)_{\alpha\beta}
 \tr \Bigl\{ \tilde{\psi}_{\alpha , -r} \Bigl(
[ \tilde{X}_{i},\tilde{\psi}_{\beta}] \Bigr)_r \Bigr\} \Biggr] \ .
\label{bfss_action_cutoff}
\eeqa
%
It is important that we have introduced the 
cutoff $\Lambda$ after fixing the gauge 
{\em non-perturbatively}.
This is possible only in 1d. In higher dimensions,
the momentum cutoff regularization inevitably 
breaks the gauge invariance.
In the bosonic case, we have checked explicitly
\cite{Hanada-Nishimura-Takeuchi} that
the results of the {\em non-lattice} 
simulation agree
with the results of the lattice simulation
in the continuum limit.

Note
that our action is nothing but the
gauge-fixed action in the continuum
except for having a Fourier mode cutoff.
This leads to various advantages
over the lattice approach proposed 
in ref.\ \cite{Catterall:2007fp}.
Supersymmetry, which is mildly broken by
the cutoff, is shown 
(in 1d Wess-Zumino model)
to be restored much faster 
than the continuum limit is achieved.
In fact, the continuum limit is also
approached faster than one would naively expect
from the number of degrees of freedom.
These are understandable from the fact that
the modes above the cutoff are naturally
suppressed by the kinetic term.
A further
(albeit technical) 
advantage of our formulation
is that the Fourier acceleration,
which eliminates the 
critical slowing down completely 
\cite{Catterall:2001jg},
can be implemented {\em without extra cost}
since we are dealing with Fourier modes directly.
We consider that all 
these merits of the present
approach compensate the superficial
increase in the computational effort 
by the factor of O($\Lambda$)
compared to the lattice 
approach \cite{Catterall:2007fp}
with the same number of degrees of freedom.

The fermionic action $S_{\rm f}$ may be
written in the form
$S_{\rm f}
= \frac{1}{2} {\cal M}_{A \alpha r ; B \beta s}
\tilde{\psi}_{\alpha r}^A \tilde{\psi}_{\beta s}^B
$,
where we have expanded
$\tilde{\psi}_{\alpha r}
= \sum_{A=1}^{N^2} \tilde{\psi}_{\alpha r}^A t^A$
in terms of U($N$) generators $t^A$.
Integrating out the fermions,
we obtain
the Pfaffian ${\rm Pf}{\cal M}$, which is complex
for generic configurations of the remaining bosonic variables.
However, 
it turns out to be
real positive with high accuracy in the temperature
region studied in the present work.
Hence we can replace it by 
$|{\rm Pf}{\cal M}|
= {\rm det} ( {\cal D}^{1/4})$,
where ${\cal D}={\cal M}^\dag {\cal M}$.
One can then apply the Rational Hybrid Monte Carlo
algorithm \cite{Clark:2004cp} to study the system
in an efficient way.

\section{Results}
 
\begin{figure}[htb]
\begin{center}
\includegraphics[height=6cm]{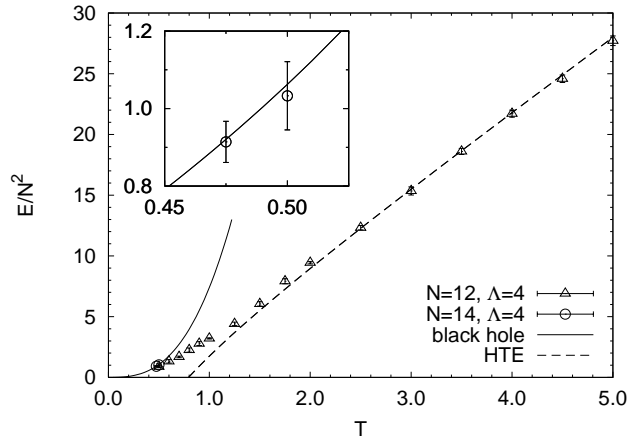}
\end{center}
\caption{
The energy
(normalized by $N^2$)
is plotted against $T$.
The dashed line represents the result
obtained by HTE
up to the next leading order 
for $N=12$ \cite{HTE}.
The solid line represents
the asymptotic
power-law behavior at small $T$
predicted by the gauge/gravity duality.
The upper left panel zooms
up the region, where
the power-law behavior sets in.
}
\label{energy}
\end{figure}

In fig.\ \ref{energy}
we plot the internal energy defined by
$E = \frac{\del}{\del \beta } (\beta {\cal F})$,
where ${\cal F}$ is the free energy
of the system.
Our results interpolate nicely the
weak coupling behavior --- calculated
by the high temperature expansion (HTE) 
up to the
next leading order \cite{HTE} ---
and the strong coupling behavior 
$\frac{E}{N^2}
= 7.4 \cdot T^{2.8}$
predicted by the gauge/gravity duality 
\cite{Itzhaki:1998dd} from
the dual black-hole geometry \cite{Klebanov:1996un}.
The power-law behavior sets in
at $T \simeq 0.5$, 
which is reasonable since
the effective coupling constant
is given by
$\tilde{\lambda}=1/T^3$ in our convention.

In ref.\ \cite{KLL}
the Gaussian expansion method 
was applied to the present model,
and the energy 
obtained at the leading order
was fitted nicely to the power law
$E/N^2 = 3.4 \cdot T^{2.7}$ within
$0.25 \lesssim T \lesssim 1$.
This result is in reasonable agreement
with our data at $T \sim 1$, but
disagrees at lower temperature.

\begin{figure}[htb]
\begin{center}
\includegraphics[height=6cm]{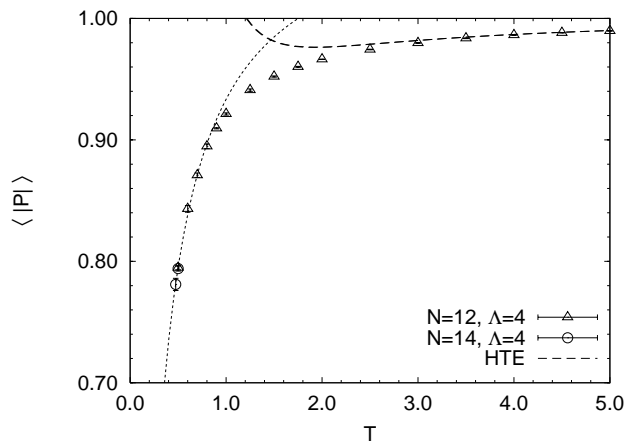}
\end{center}
\caption{
The Polyakov line
is plotted against 
$T$.
The dashed line represents the result
of HTE
up to the next leading order for $N=12$ \cite{HTE}.
The dotted line represents a fit to
eq.\ (\protect\ref{polya-fit}) 
with $a=0.15$ and $b=0.072$.
}
\label{polyakov}
\end{figure}

\begin{figure}[htb]
\begin{center}
\includegraphics[height=6cm]{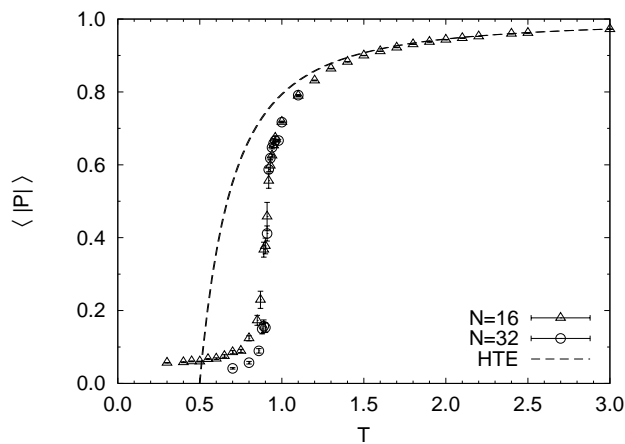}
\end{center}
\caption{
The Polyakov line
for the bosonic model
\cite{Kawahara:2007fn}.
The dashed line represents
the result of HTE
up to the next leading order 
for $N=16$ \cite{HTE}.
}
\label{polyakov_bos}
\end{figure}

In fig.\ \ref{polyakov}
we plot the absolute value 
of the Polyakov line
$P = \frac{1}{N}\sum_{a=1}^N \ee^{i \alpha_a}$, 
which is the
order parameter for the SSB of
the U(1) symmetry.
It changes smoothly 
for the range
of $T$ investigated,
which implies the absence of a phase transition
as predicted by the gauge/gravity 
duality 
\cite{Barbon:1998cr,Aharony4}.
At low $T$ it
can be fitted nicely to the asymptotic behavior
characteristic to
a deconfined theory:
\beq
\langle |P| \rangle = \exp
\left(-\frac{a}{T}+b \right) \ .
\label{polya-fit}
\eeq
This is in striking contrast to
the bosonic case
\cite{Kawahara:2007fn}
shown in fig.\ \ref{polyakov_bos} 
for comparison\footnote{In 
ref.\ \cite{Kawahara:2007fn} it was found that
there are actually three phases in the bosonic model.
The intermediate phase appears in a
very narrow range of temperature 
$T_{{\rm c}2} < T < T_{{\rm c}1}$, where 
$ T_{{\rm c}1}= 0.905(2)$ and $T_{{\rm c}2}=0.8761(3)$,
and it is characterized by the non-uniform
eigenvalue distribution of the holonomy matrix.
The order of phase transitions are 
second order at $ T= T_{{\rm c}1}$,
and third order at $ T=T_{{\rm c}2}$.}.


\section{Summary and future prospects}
We have presented
the first Monte Carlo results
for the maximally supersymmetric matrix
quantum mechanics.
%
The non-lattice simulation
enabled us to 
study the low temperature 
behavior, which was not
accessible by HTE.
This provided highly non-trivial evidences
for
the gauge/gravity duality.
In particular, we observed that
the internal energy
asymptotes nicely at low temperature
to the result 
obtained from the dual black-hole 
geometry.

Our results suggest that
not only the power but also 
the coefficient
of the power-law behavior 
is
reproduced correctly by
the gauge theory in the $N\rightarrow \infty$
and $\tilde{\lambda}\rightarrow \infty$ limits.
This implies that
we were able to identify
the microscopic degrees of freedom,
which accounts for
the Bekenstein-Hawking entropy 
for the 10d non-extremal black hole.
They are nothing but 
the open strings attached to the
D0-branes, which are described 
by the gauge theory.
This should be compared with
ref.\ \cite{Strominger:1996sh},
which studied extremal black holes
and relied on the supersymmetric 
non-renormalization theorem.

Assuming the duality to hold
in the stronger sense, one may
go on and investigate the quantum and stringy 
corrections to
the black-hole 
thermodynamics
from the gauge theory side as finite-$N$ and 
finite-$\tilde{\lambda}$ effects. 
In particular 
it would be interesting to understand
the physical meaning of the
infrared instability 
observed in our simulation \cite{AHNT} 
from that perspective.

When we simulate M theory,
we should impose periodic boundary conditions
on fermions, and then
the system with finite $N$ corresponds to 
a sector of M theory compactified on a light-like 
circle \cite{Seiberg-Sen}.
%
However, the Pfaffian will not be close to 
real positive due to the fermionic zero modes
unlike the situation in the present work.
That may cause a technical problem
known as
the sign problem when one tries to 
investigate the large-$N$ behavior.
The same problem occurs in Monte Carlo studies
of the totally reduced models,
in which the phase of the Pfaffian is speculated 
\cite{NV} to
induce the spontaneous breaking of SO(10)
symmetry down to SO(4).
This pattern of SSB is indeed suggested
by the Gaussian expansion method \cite{SSB},
and it provides a natural scenario for the
dynamical generation of 4d space-time \cite{AIKKT}. 
In ref.\ \cite{Kawahara:2005an} it was conjectured,
based on the Eguchi-Kawai equivalence,
that a similar phenomenon occurs also in the
supersymmetric matrix quantum mechanics.
We hope to address such an issue
from first principles
by using the non-lattice simulation method
together with the idea proposed in
ref.\ \cite{Anagnostopoulos:2001yb} to
overcome the sign problem.



\end{document}